\documentclass[11pt,a4paper]{article}

\usepackage[T1]{fontenc}
\usepackage[utf8]{inputenc}
\usepackage[a4paper,margin=25mm]{geometry}
\usepackage{amsmath,amssymb}
\usepackage{graphicx}
\usepackage{float}
\usepackage{subfig}
\usepackage{booktabs}
\usepackage{multirow}
\usepackage{cite}
\usepackage{placeins}
\usepackage{setspace}
\usepackage[hidelinks]{hyperref}
\title{Joint Effects of Node Density, Propagation, Wi-Fi Generation, and Transport Protocol on WLAN Performance: An ns-3 Study}

\author{Leonel~Olímpio~Silima\\
\normalsize Faculty of Engineering -- University of Porto\\
\normalsize LIAAD | INESC TEC, Porto, Portugal}
\date{September 2026}

\begin{document}

\maketitle

\begin{abstract}
Wireless local area network performance is jointly shaped by medium contention, propagation, Wi-Fi configuration, and transport-layer behaviour. This study evaluates IEEE 802.11g and IEEE 802.11ax in a single-access-point uplink topology, comparing UDP and TCP under Friis, LogDistance, and cascaded Friis--Nakagami propagation. The factorial design spans 5--50 stations, six access-point distances from 5 to 50\,m, and three random seeds. Of 1,080 scheduled executions, 1,073 produced valid CSV outputs; seven runs terminated with internal ns-3/cppyy segmentation errors and were excluded. Across the global scenario summaries, TCP FlowMonitor packet-delivery ratios (PDRs) range from 0.9894 to 0.9978, whereas UDP PDRs range from 0.2978 to 0.7311. The highest mean aggregate throughput is 26.5189\,Mb/s for TCP over the configured 802.11ax mode with Friis propagation; UDP reaches 25.8845\,Mb/s in the same combination. These results show why reliability, capacity, delay, and fairness must be considered together. FlowMonitor metrics are IP-flow observations: the study does not establish application-level delivery or isolate optional 802.11ax multi-user mechanisms.
\end{abstract}

\par\medskip

\section{Introduction}
Wireless local area networks (WLANs) are used in enterprise, residential, and industrial settings where a shared radio channel must serve multiple stations. This study uses the ns-3 discrete-event network simulator for packet-level evaluation \cite{riley2010ns3}. Performance is shaped by both link conditions and medium access: path loss and fading affect frame reception, while contention and retransmissions consume airtime and can reduce the service available to other stations. Classical analyses of the IEEE 802.11 distributed coordination function show how increasing the number of contenders changes throughput and access behaviour \cite{bianchi2000}. Later Wi-Fi generations add PHY and MAC capabilities, but their realized benefit depends on which features are configured and on the surrounding channel conditions \cite{bellalta2016,mozaffariahrar2022}.

This study compares configured IEEE 802.11g and IEEE 802.11ax modes across a controlled set of station counts, distances, propagation models, and transport protocols. IEEE 802.11g provides a legacy WLAN reference \cite{vassis2005}; 802.11ax is designed to improve efficiency in dense WLANs, including through mechanisms such as OFDMA and multi-user operation \cite{bellalta2016,mozaffariahrar2022}. Advanced OFDMA scheduling, explicit MU-MIMO/spatial-stream operation, and channel-width sweeps were not configured. Accordingly, the results describe the simulated 802.11g/802.11ax configurations, not the full potential of every 802.11ax feature.

Transport choice changes what a packet-delivery metric represents. UDP does not provide transport-layer retransmission, while TCP uses feedback and congestion control to recover from loss and adapt its sending rate \cite{xylomenos1999,bruno2008,mathis1997}. Thus, a high observed TCP PDR can coexist with lower throughput, greater protocol overhead, or unfair allocation; it does not mean that the radio channel itself became loss-free. Empirical and analytical studies have previously shown that TCP and UDP interact differently with shared 802.11 airtime \cite{xylomenos1999,bruno2008}.

The work is organized around three questions:
\begin{itemize}
  \item \textbf{RQ1:} How do station density, access-point distance, and propagation model jointly relate to observed reliability, throughput, delay, and fairness?
  \item \textbf{RQ2:} How do the reported UDP and TCP outcomes differ under matched PHY, propagation, density, and distance conditions?
  \item \textbf{RQ3:} Does the configured 802.11ax mode outperform the configured 802.11g mode across all propagation and transport combinations?
\end{itemize}

This study provides a multi-metric account of the 1,080-run design, records incomplete executions, defines the FlowMonitor-based measures, and compares the resulting scenario summaries. The analysis uses scenario-level CSV summaries and generated figures; it does not include new raw-trace analysis or application-level validation.

\section{Background and Related Work}
\subsection{WLAN contention and Wi-Fi generations}
In infrastructure WLANs, stations share access to the channel through contention-based medium access. Increasing the number of active stations can raise aggregate offered traffic while also increasing contention; aggregate throughput may therefore rise initially and then saturate, while per-station throughput and fairness decline. Bianchi's widely used analytical model characterizes throughput behaviour under the IEEE 802.11 distributed coordination function \cite{bianchi2000}. IEEE 802.11g extends high-rate WLAN operation in the 2.4\,GHz band \cite{vassis2005}. IEEE 802.11ax targets high-efficiency operation, particularly in denser networks, using PHY/MAC improvements that include multi-user mechanisms \cite{bellalta2016,mozaffariahrar2022}. The present experiment does not isolate those optional mechanisms and should not be read as a feature-by-feature 802.11ax benchmark.

\subsection{Propagation models}
The Friis relation represents ideal free-space propagation and is useful as an optimistic line-of-sight reference \cite{friis1946}. The LogDistance model extends a reference path loss by a distance-dependent exponent; path-loss exponent selection and fitting are environment-dependent \cite{karttunen2016}. Nakagami-$m$ fading models a distribution of received envelope amplitude and can represent varying fading severity \cite{nakagami1960}. The third propagation condition combines deterministic distance attenuation with small-scale fading through a cascaded Friis--Nakagami model. The model parameter values were not documented. The equations below explain the physical quantities but are not used to reconstruct or modify the simulation outcomes.

\subsection{Transport protocols and measurement}
Comparisons of TCP and UDP over wireless LANs have long shown that differences in reliability and throughput depend on the interaction between transport control and the shared wireless medium \cite{xylomenos1999,bruno2008}. Flow-level measurements were collected with ns-3 FlowMonitor, which records packet, byte, loss, timing, and delay statistics \cite{carneiro2010flowmonitor}. FlowMonitor measurements are made at the IP-flow level. They are therefore not equivalent to application goodput, unique TCP sequence-number delivery, or PHY/MAC-layer error statistics.

\section{Experimental Design}
\subsection{Scenario matrix}
The topology contains one access point (AP) and multiple static wireless stations (STAs), with application traffic flowing from STA to AP. UDP and TCP were run as separate campaigns. Each campaign crossed two Wi-Fi modes, three propagation configurations, five station counts, six AP-distance settings, and three random seeds. The documented settings are summarized in Table~\ref{tab:design}.

\begin{table}[!htbp]
\caption{Documented Simulation Factors and Traffic Settings}
\label{tab:design}
\centering
\footnotesize
\begin{tabular}{p{0.22\textwidth}p{0.69\textwidth}}
\toprule
\textbf{Factor} & \textbf{Documented setting}\\
\midrule
Topology and direction & One AP; static STAs; uplink application traffic (STA $\rightarrow$ AP).\\
Wi-Fi modes & IEEE 802.11g and IEEE 802.11ax modes in ns-3.\\
Transport & UDP and TCP, executed as independent campaigns.\\
Propagation & Friis; LogDistance; cascaded Friis--Nakagami.\\
Station count, $N_s$ & 5, 10, 20, 30, and 50.\\
AP-distance levels & 5, 10, 20, 30, 40, and 50\,m.\\
Randomness & Three scheduled seeds/run indices: 1, 2, and 3.\\
Simulation/application timing & 10\,s simulation; sinks start at 1\,s and sources at approximately 2\,s.\\
Application packet & 1024-byte payload.\\
Per-source offered rate, $r_s$ & 1\,Mb/s for 5 stations; 2\,Mb/s for 10--50 stations.\\
Rate control & \texttt{IdealWifiManager}.\\
\bottomrule
\end{tabular}
\end{table}

For each transport campaign, the scheduled run count was
\begin{equation}
N_{\mathrm{runs}} =
N_{\mathrm{PHY}}N_{\mathrm{prop}}N_{\mathrm{STA}}N_{\mathrm{dist}}N_{\mathrm{seed}}
=2\cdot3\cdot5\cdot6\cdot3=540.
\label{eq:runs}
\end{equation}
Thus, the complete design scheduled 1,080 executions. Removing the seed factor leaves $2\cdot3\cdot5\cdot6=180$ grouped scenario cells per transport and 360 cells in total.

\subsection{Execution completeness and processing}
The campaign yielded 536 valid UDP outputs and 537 valid TCP outputs. Four UDP and three TCP executions terminated with internal ns-3/cppyy segmentation failures. The seven failed executions were excluded from aggregation, leaving 1,073 valid outputs. Since the intended replication count was three, scenario groups affected by a failure have fewer available replications. Scenario-level aggregates retain the number of available runs and the associated confidence-interval fields.

Randomness was controlled with the ns-3 RngSeedManager and three independent run indices. The processing pipeline generated the experimental matrices, ran ns-3, exported per-flow and aggregate CSVs, summarized outputs across seeds, aligned equivalent UDP/TCP scenario cells, and produced figures and summary tables. In TCP post-processing, reverse-direction ACK flows were retained in raw outputs but excluded from useful-throughput aggregates so that the aggregate represented STA-to-AP data flows. The available research materials contain aggregate comparison CSVs and final figures, but not run-level raw CSVs or the simulation and analysis scripts. Independent reruns therefore require recovery of these materials.

\subsection{Propagation and link quantities}
The following equations clarify the physical interpretation by linking distance and fading to received power and packet errors. This campaign does not provide measured SNR, bit-error rate, or PHY packet-error probability.

Carrier wavelength at frequency $f_c$ is
\begin{equation}
\lambda = \frac{c}{f_c},
\label{eq:wavelength}
\end{equation}
where $c$ is the speed of light. Under ideal free-space line-of-sight propagation, the Friis received power is
\begin{equation}
P_r = P_t G_t G_r\left(\frac{\lambda}{4\pi d}\right)^2,
\label{eq:friis}
\end{equation}
where $P_t$ is transmit power, $G_t$ and $G_r$ are antenna gains, and $d$ is separation. The corresponding free-space path loss is
\begin{equation}
L_{\mathrm{FS}}(d)\,[\mathrm{dB}]
=20\log_{10}\left(\frac{4\pi d}{\lambda}\right).
\label{eq:fspl}
\end{equation}
The LogDistance path-loss relation can be written as
\begin{equation}
PL(d)=PL(d_0)+10n\log_{10}\left(\frac{d}{d_0}\right),\qquad d\ge d_0,
\label{eq:logdistance}
\end{equation}
where $d_0$ is a reference distance and $n$ is the path-loss exponent.

For a Nakagami-$m$ envelope $R$, the probability density is
\begin{equation}
f_R(r)=
\frac{2m^m}{\Gamma(m)\Omega^m}r^{2m-1}
\exp\left(-\frac{mr^2}{\Omega}\right),\quad r\ge0,
\label{eq:nakagami}
\end{equation}
where $m$ controls fading severity, $\Omega=\mathbb{E}[R^2]$, and $\Gamma(\cdot)$ is the gamma function. A conceptual signal-to-noise ratio is
\begin{equation}
\gamma=\frac{P_r}{N},
\label{eq:snr}
\end{equation}
where $N$ denotes received noise power over the modeled channel bandwidth. If bit errors were independent with probability $P_b$, an illustrative $L$-bit packet error probability would be
\begin{equation}
P_{\mathrm{e}}=1-(1-P_b)^L.
\label{eq:per}
\end{equation}
Equations~\eqref{eq:snr}--\eqref{eq:per} explain a possible physical mechanism; they are not measurements from this campaign and must not be interpreted as a calculation of its reported PDR.

\subsection{Flow-level performance metrics}
Let $B_{\mathrm{rx}}$ be the received FlowMonitor IP-byte count for the selected data flow(s), $\Delta t_f$ the observation interval used in the source analysis, and $N_{\mathrm{tx}}$ and $N_{\mathrm{rx}}$ the corresponding transmitted and received FlowMonitor packet counts. Aggregate throughput and packet delivery are
\begin{equation}
\begin{aligned}
T_{\mathrm{agg}}&=\frac{8B_{\mathrm{rx}}}{\Delta t_f}\quad[\mathrm{bit/s}],\\
\mathrm{PDR}&=\frac{N_{\mathrm{rx}}}{N_{\mathrm{tx}}},\qquad
\mathrm{PLR}=1-\mathrm{PDR}.
\end{aligned}
\label{eq:flowmetrics}
\end{equation}
Let $D_{\mathrm{sum}}$ denote the accumulated FlowMonitor delay, i.e., the sum of per-packet delays for received packets. The reported mean delay is
\begin{equation}
\bar{D}=\frac{D_{\mathrm{sum}}}{N_{\mathrm{rx}}}.
\label{eq:delay}
\end{equation}
If $N_s$ stations each offer rate $r_s$, the nominal application offered load, per-station throughput, and reported relative efficiency are
\begin{equation}
L_{\mathrm{off}}=N_s r_s,\qquad
T_{\mathrm{node}}=\frac{T_{\mathrm{agg}}}{N_s},\qquad
\eta_{\mathrm{rel}}=\frac{T_{\mathrm{agg}}}{L_{\mathrm{off}}}.
\label{eq:load}
\end{equation}
The last quantity is a throughput-to-configured-load ratio, not PHY spectral efficiency. The report notes that values may exceed one because FlowMonitor counts IP bytes over its observation interval while the denominator is the nominal application payload rate.

For per-flow throughputs $x_i$ across $N_f$ station flows, Jain's fairness index is
\begin{equation}
J=\frac{\left(\sum_{i=1}^{N_f}x_i\right)^2}
{N_f\sum_{i=1}^{N_f}x_i^2}.
\label{eq:jain}
\end{equation}
Values nearer one indicate a more even flow-throughput allocation; lower values indicate greater disparity \cite{sediq2013}.

\subsection{Seed-level summary and uncertainty}
For a metric $X$ observed over $R$ valid independent seeds, the sample mean and sample variance are
\begin{equation}
\bar{X}=\frac{1}{R}\sum_{r=1}^{R}X_r,\qquad
s^2=\frac{1}{R-1}\sum_{r=1}^{R}(X_r-\bar{X})^2.
\label{eq:sample}
\end{equation}
The conventional two-sided 95\% Student-$t$ interval is
\begin{equation}
\bar{X}\ \pm\ t_{0.975,R-1}\frac{s}{\sqrt{R}}.
\label{eq:ci}
\end{equation}
Independent replications are a standard basis for simulation confidence intervals, but the precision of such intervals is limited when the number of replications is small \cite{law1977}. Here, three seeds were scheduled and some groups have only two valid runs. Scenario-level CSV outputs include per-cell counts and \texttt{ci95} fields. The global tables report the calculated global means; no additional confidence bounds or hypothesis tests are introduced.

\section{Results}
\subsection{Global scenario means}
Tables~\ref{tab:capacity} and~\ref{tab:service} summarize the 12 protocol--standard--propagation combinations. Throughput and per-station throughput are in Mb/s; delays are in seconds. Delay precision is retained sufficiently to avoid displaying a small positive value as zero.

\begin{table}[!htbp]
\caption{Reported Global Capacity and Throughput-to-Load Summaries}
\label{tab:capacity}
\centering
\scriptsize
\setlength{\tabcolsep}{4pt}
\begin{tabular}{lllrrr}
\toprule
\textbf{Transport} & \textbf{Wi-Fi mode} & \textbf{Propagation} &
\textbf{Throughput} & \textbf{Per station} & \textbf{$\eta_{\mathrm{rel}}$}\\
& & & \textbf{(Mb/s)} & \textbf{(Mb/s)} & \\
\midrule
TCP & 802.11ax & Friis & 26.5189 & 1.3501 & 0.7850\\
TCP & 802.11ax & Friis--Nakagami & 17.9799 & 1.0473 & 0.6320\\
TCP & 802.11ax & LogDistance & 8.1914 & 0.5170 & 0.3282\\
TCP & 802.11g & Friis & 21.0817 & 1.4859 & 0.9551\\
TCP & 802.11g & Friis--Nakagami & 11.0286 & 0.9762 & 0.6946\\
TCP & 802.11g & LogDistance & 10.6835 & 0.8262 & 0.5515\\
UDP & 802.11ax & Friis & 25.8845 & 1.3031 & 0.7544\\
UDP & 802.11ax & Friis--Nakagami & 21.0900 & 1.1372 & 0.6714\\
UDP & 802.11ax & LogDistance & 7.7080 & 0.4856 & 0.3079\\
UDP & 802.11g & Friis & 17.1129 & 1.0232 & 0.6145\\
UDP & 802.11g & Friis--Nakagami & 10.1943 & 0.6963 & 0.4489\\
UDP & 802.11g & LogDistance & 7.6043 & 0.4820 & 0.3061\\
\bottomrule
\end{tabular}
\end{table}

\begin{table}[!htbp]
\caption{Reported Global Reliability, Delay, and Fairness Summaries}
\label{tab:service}
\centering
\scriptsize
\setlength{\tabcolsep}{5pt}
\begin{tabular}{lllrrr}
\toprule
\textbf{Transport} & \textbf{Wi-Fi mode} & \textbf{Propagation} &
\textbf{PDR} & \textbf{Mean delay (s)} & \textbf{Jain fairness}\\
\midrule
TCP & 802.11ax & Friis & 0.9978 & 0.006802 & 0.7475\\
TCP & 802.11ax & Friis--Nakagami & 0.9928 & 0.025460 & 0.6562\\
TCP & 802.11ax & LogDistance & 0.9963 & 0.000190 & 0.2994\\
TCP & 802.11g & Friis & 0.9953 & 0.009628 & 0.7334\\
TCP & 802.11g & Friis--Nakagami & 0.9894 & 0.040119 & 0.4906\\
TCP & 802.11g & LogDistance & 0.9957 & 0.000840 & 0.2971\\
UDP & 802.11ax & Friis & 0.7311 & 0.017098 & 0.7504\\
UDP & 802.11ax & Friis--Nakagami & 0.6423 & 0.068773 & 0.6843\\
UDP & 802.11ax & LogDistance & 0.2997 & 0.000049 & 0.2996\\
UDP & 802.11g & Friis & 0.5971 & 0.121485 & 0.6926\\
UDP & 802.11g & Friis--Nakagami & 0.4359 & 0.159940 & 0.4897\\
UDP & 802.11g & LogDistance & 0.2978 & 0.004719 & 0.2994\\
\bottomrule
\end{tabular}
\end{table}

The highest global mean throughput occurs for TCP with configured 802.11ax under Friis propagation (26.5189\,Mb/s); UDP in the same configuration reaches 25.8845\,Mb/s. Under Friis and 802.11g, TCP and UDP means are 21.0817 and 17.1129\,Mb/s, respectively. The ranking is not universal: under 802.11ax Friis--Nakagami, the UDP mean (21.0900\,Mb/s) exceeds the TCP mean (17.9799\,Mb/s). The results therefore do not support a protocol-only or standard-only ranking.

The global TCP PDR means span 0.9894--0.9978, while UDP means span 0.2978--0.7311. In both LogDistance combinations, TCP has a PDR near 0.996 but mean aggregate throughput is 8.1914\,Mb/s for 802.11ax and 10.6835\,Mb/s for 802.11g; Jain fairness is approximately 0.30. This combination of metrics is consistent with TCP recovery/adaptation coexisting with restricted capacity and unequal flow service. Because the counter is from FlowMonitor at IP-flow level, it is not proof of unique application payload delivery.

\subsection{Protocol-level distributions}
\begin{figure}[!htbp]
\centering
\subfloat[Aggregate throughput (Mb/s).]{\includegraphics[width=0.475\textwidth]{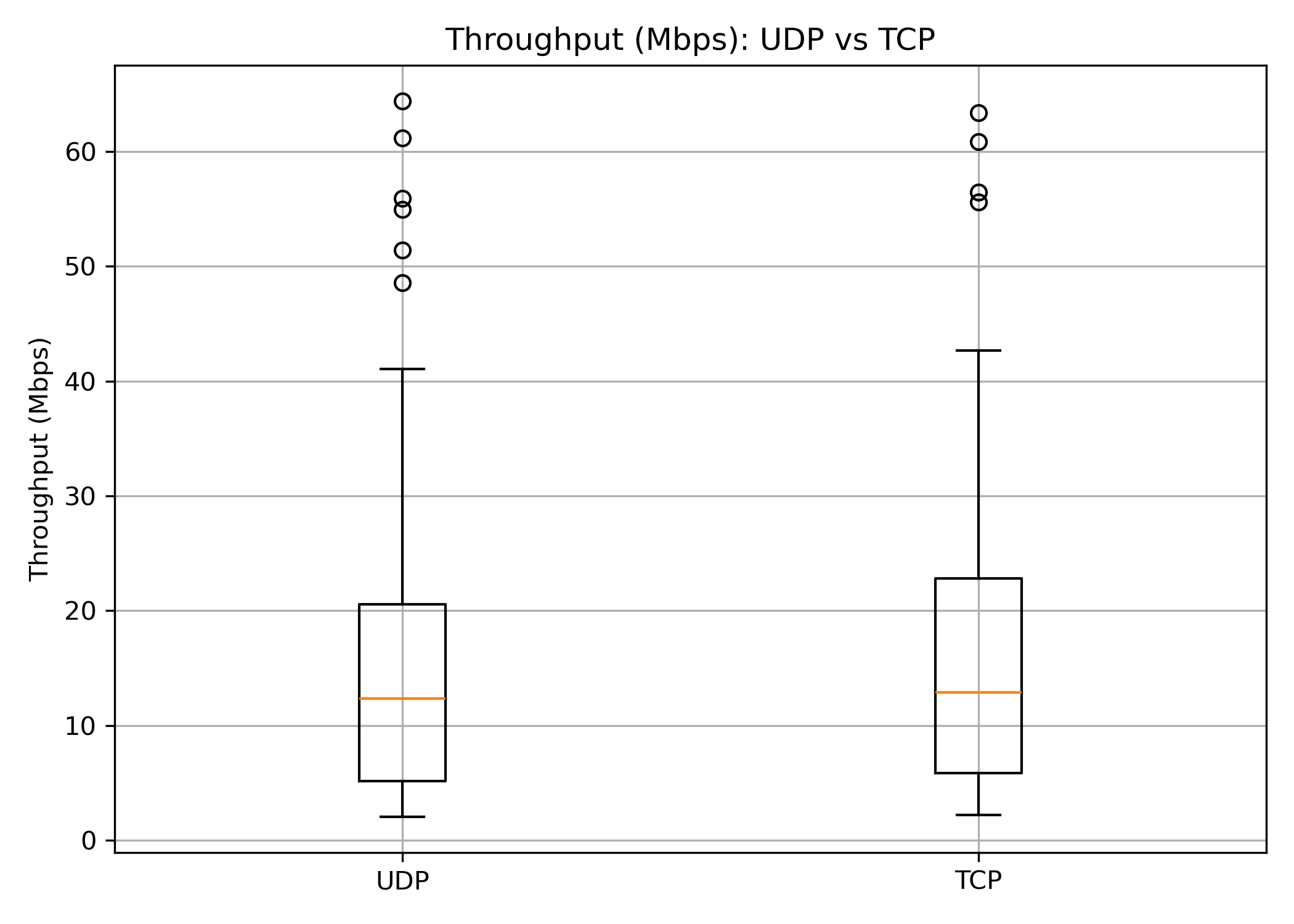}\label{fig:box-throughput}}
\hfill
\subfloat[FlowMonitor PDR.]{\includegraphics[width=0.475\textwidth]{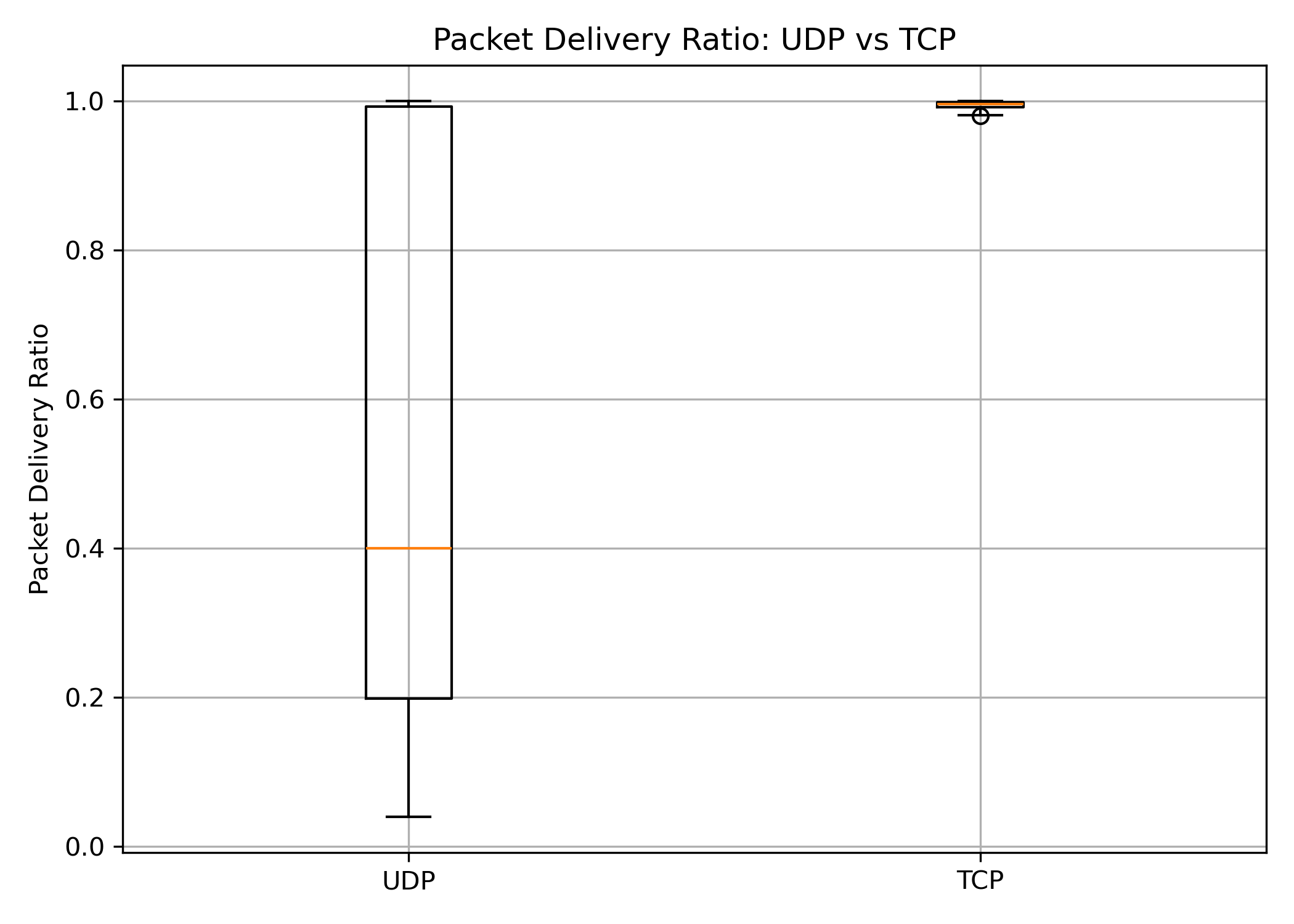}\label{fig:box-pdr}}
\caption{Protocol-level distributions across the grouped scenario outputs.}
\label{fig:boxplots1}
\end{figure}

Figure~\ref{fig:boxplots1} compares aggregate throughput and FlowMonitor PDR across the grouped scenario distributions. Throughput distributions overlap substantially, while the PDR distributions are separated: UDP exposes loss not recovered by the transport, whereas TCP observations cluster close to one. This comparison reflects the entire set of scenario groups and should not be mistaken for an isolated PHY-layer measurement.

\begin{figure}[!htbp]
\centering
\subfloat[Mean delay (s).]{\includegraphics[width=0.315\textwidth]{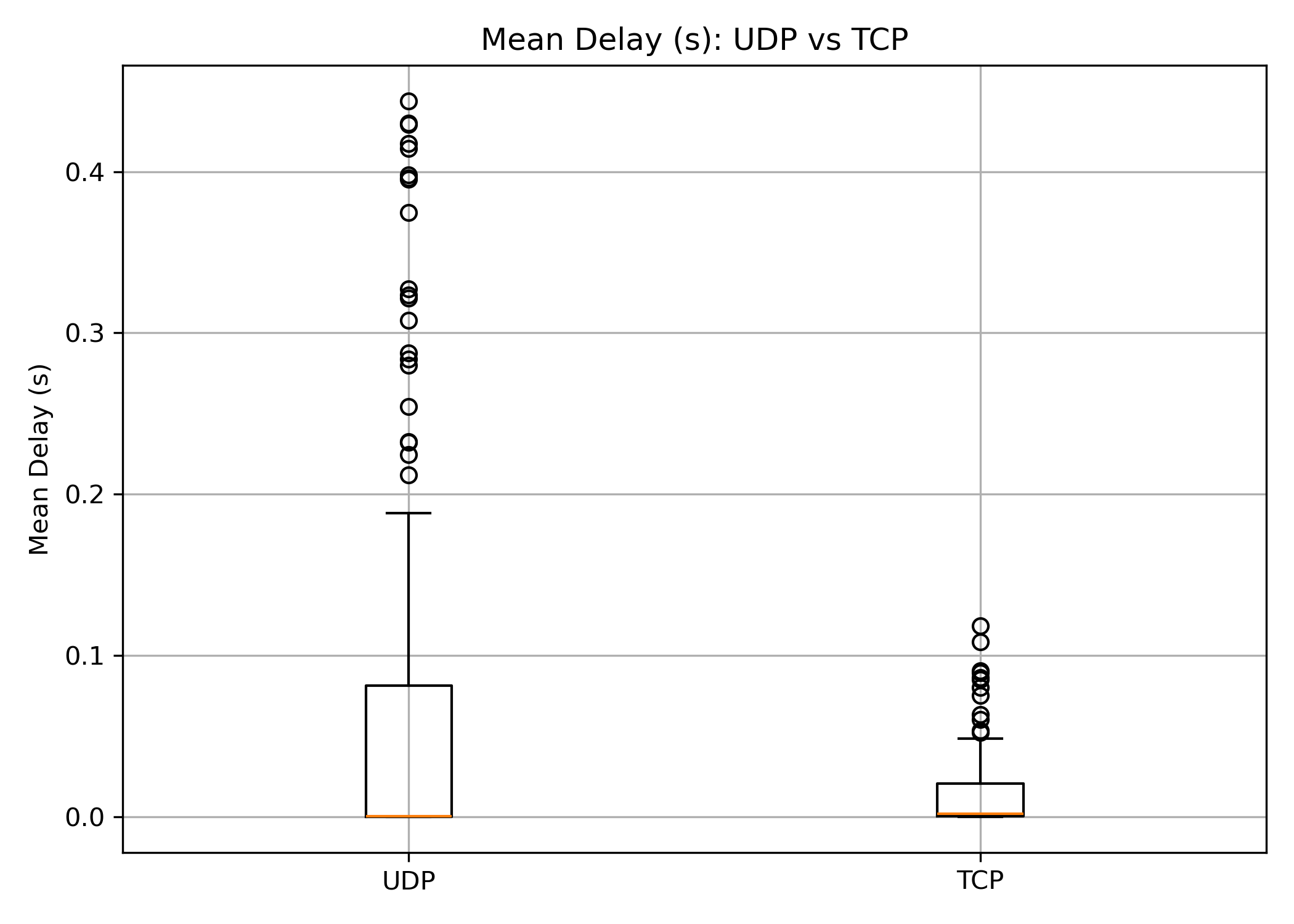}\label{fig:box-delay}}
\hfill
\subfloat[Jain fairness.]{\includegraphics[width=0.315\textwidth]{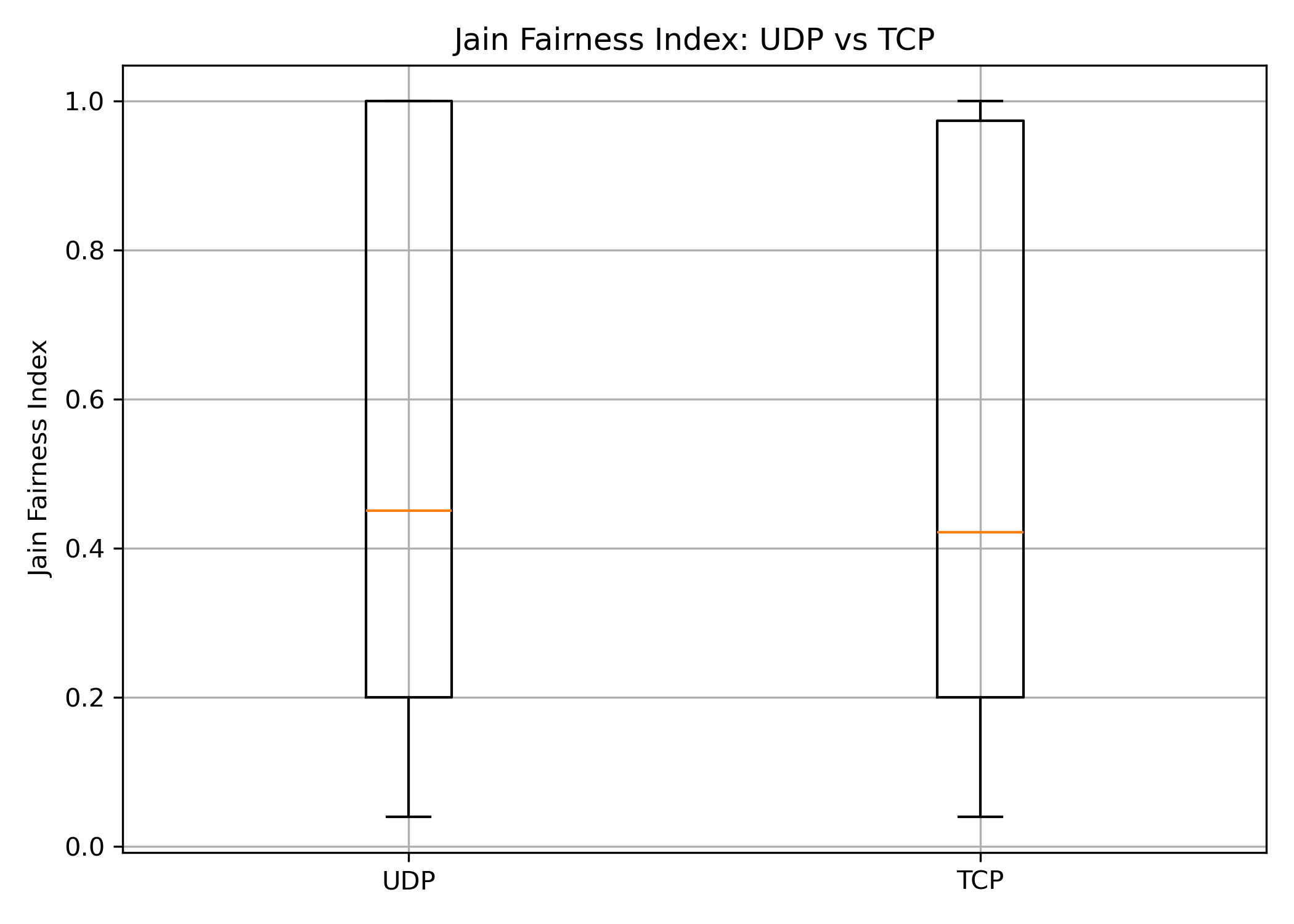}\label{fig:box-fairness}}
\hfill
\subfloat[Throughput-to-load ratio.]{\includegraphics[width=0.315\textwidth]{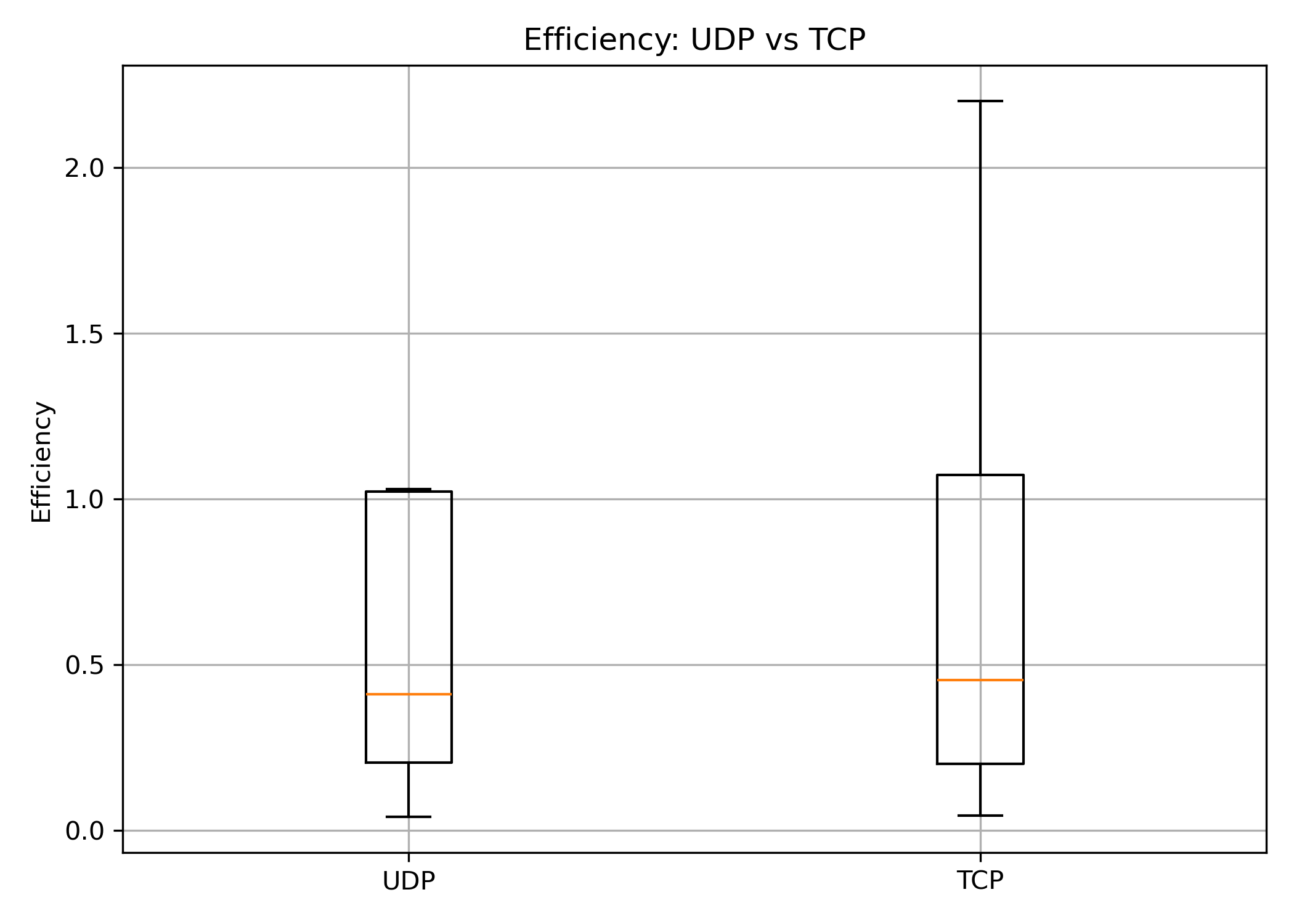}\label{fig:box-efficiency}}
\caption{Distributions of delay, fairness, and relative throughput-to-load ratio across grouped scenarios.}
\label{fig:boxplots2}
\end{figure}

Figure~\ref{fig:boxplots2} combines the remaining metric distributions. The delay distribution is lower for TCP in these processed results, but that observation is not an unconditional latency guarantee: transport response and the subset of packets that arrive affect the measured mean. Fairness is heterogeneous across scenario groups. The efficiency ratio can exceed one in individual groups because its numerator is based on IP bytes and an observed flow interval while its denominator is the nominal application payload rate; it is comparative, not a physical-layer efficiency.

\subsection{Reliability by density and distance}
The heatmaps show how PDR changes over the density--distance grid. Under UDP, PDR falls markedly in the long-distance, high-density part of the 802.11ax LogDistance grid, reaching approximately 0.04 in the 50-station, long-distance region. Under TCP, the same configured radio condition retains PDR near 0.99, but this transport-level difference coexists with low throughput and low fairness in the global summaries. A corresponding contrast appears for 802.11g with Friis--Nakagami. These are descriptive patterns in the scenario aggregates, not a claim that TCP changes the underlying propagation channel.

\begin{figure}[!htbp]
\centering
\subfloat[UDP, 802.11ax, LogDistance.]{\includegraphics[width=0.475\textwidth]{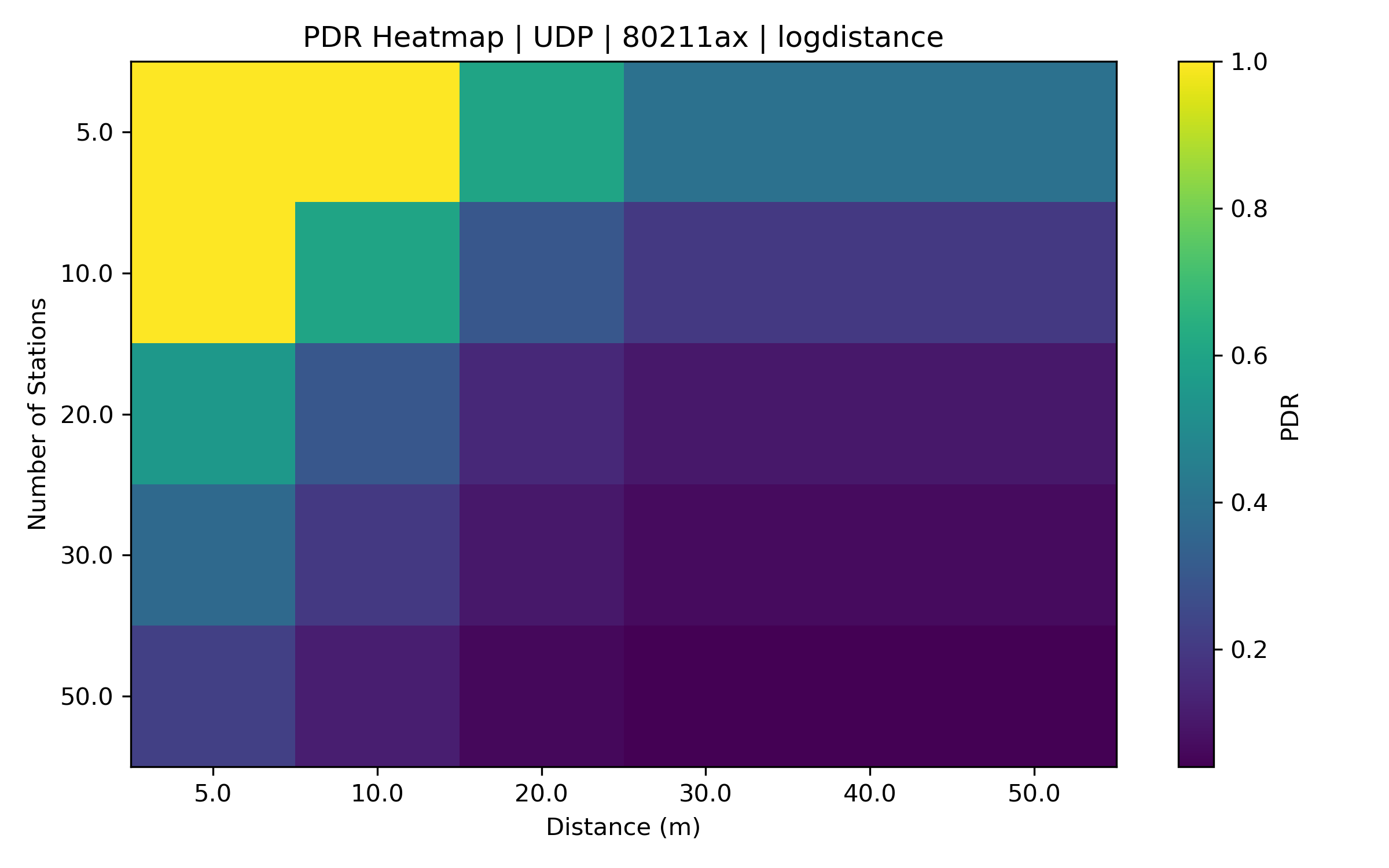}\label{fig:pdr-udp-ax-log}}
\hfill
\subfloat[TCP, 802.11ax, LogDistance.]{\includegraphics[width=0.475\textwidth]{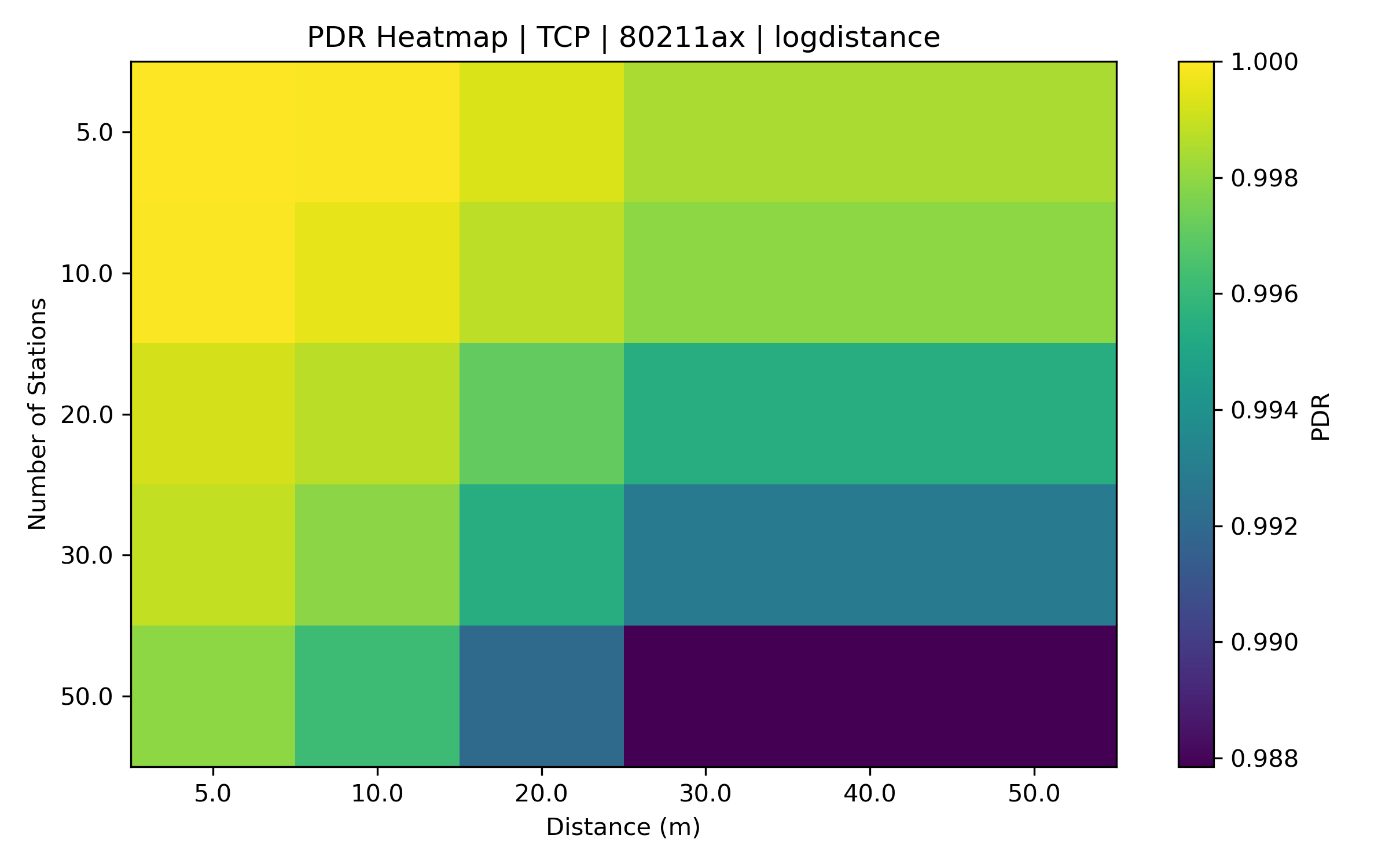}\label{fig:pdr-tcp-ax-log}}\\
\subfloat[UDP, 802.11g, Friis--Nakagami.]{\includegraphics[width=0.475\textwidth]{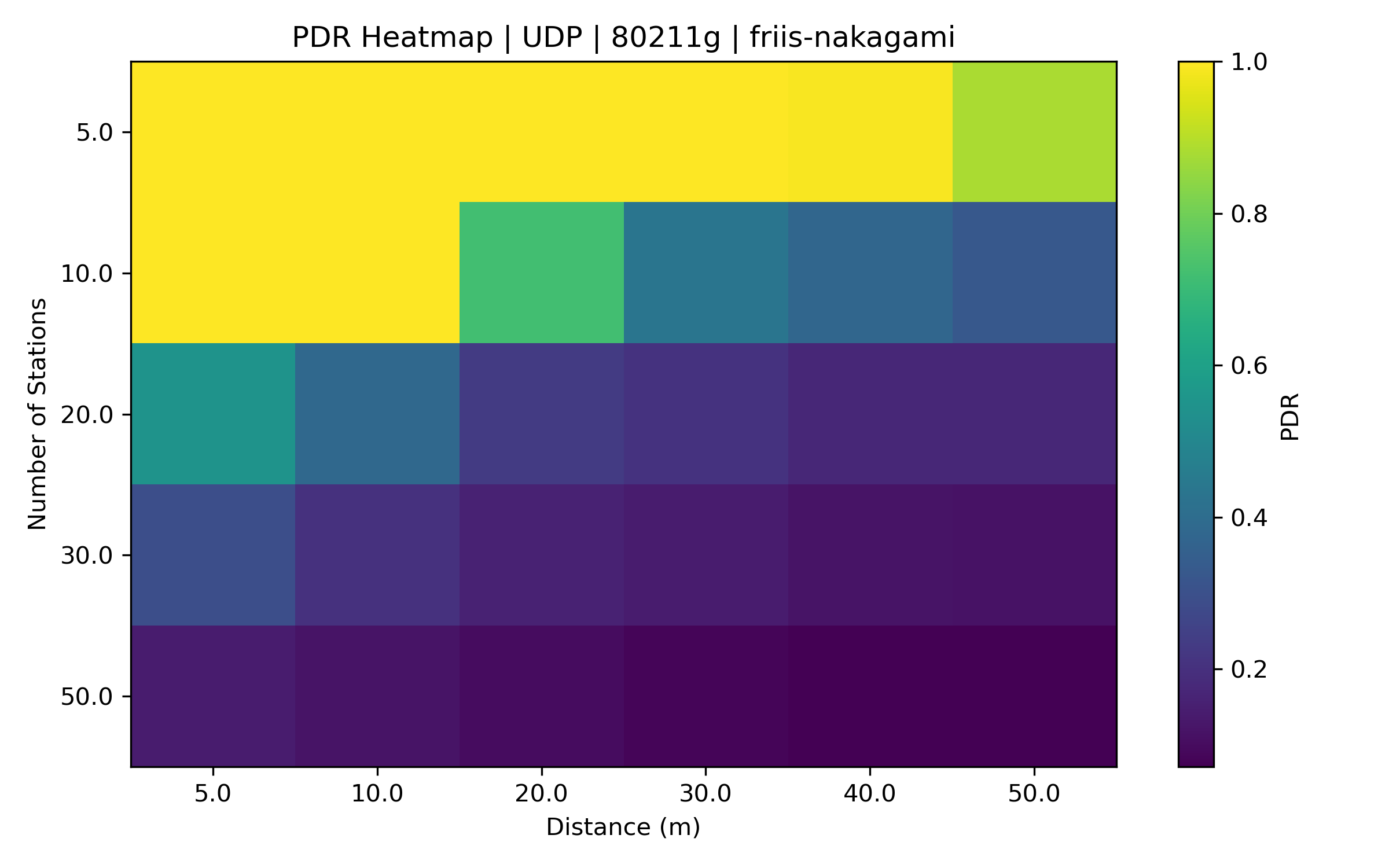}\label{fig:pdr-udp-g-nak}}
\hfill
\subfloat[TCP, 802.11g, Friis--Nakagami.]{\includegraphics[width=0.475\textwidth]{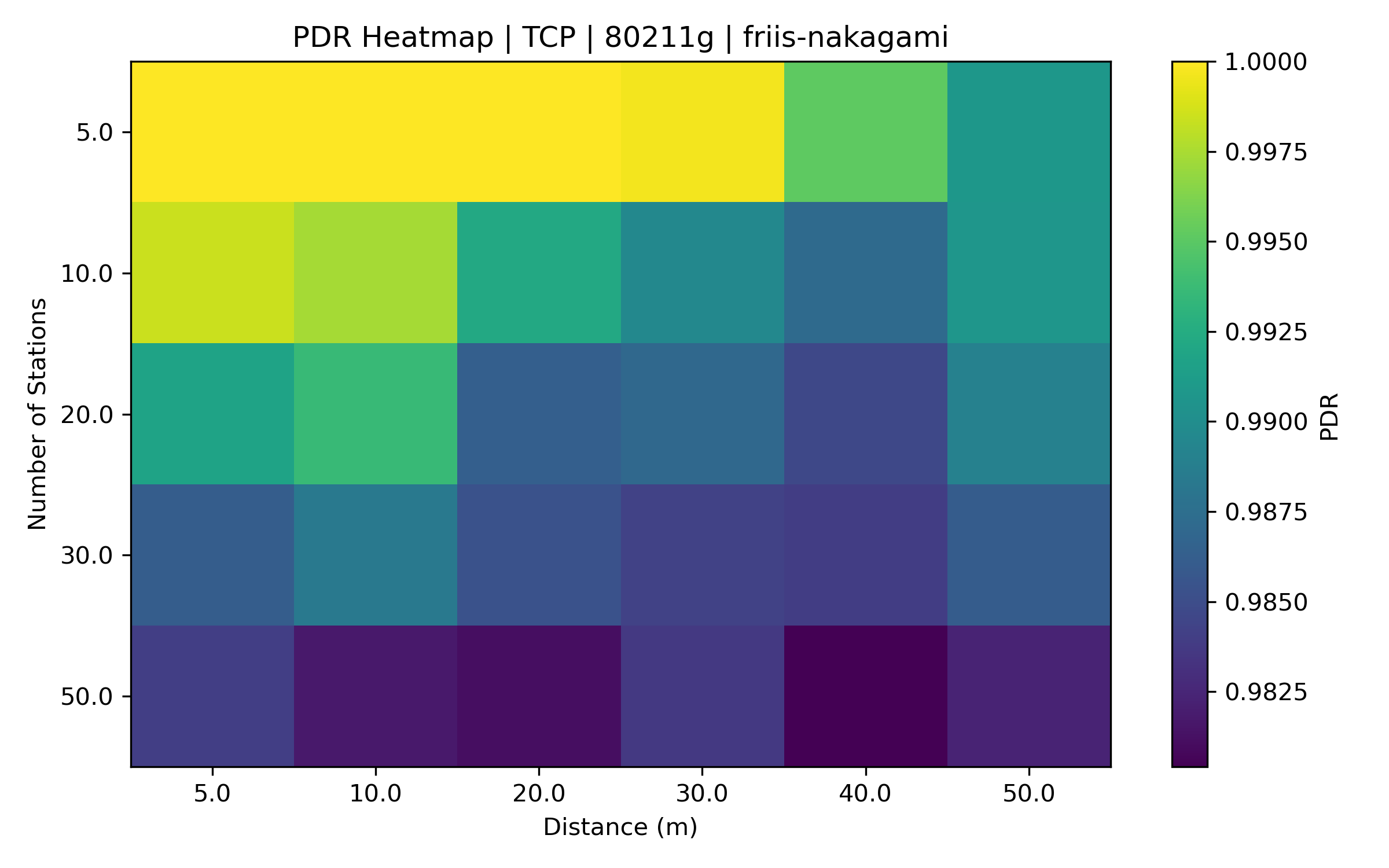}\label{fig:pdr-tcp-g-nak}}
\caption{Selected PDR heatmaps over station count and configured distance. The color scales are those of the original figures and differ between panels.}
\label{fig:pdrheatmaps}
\end{figure}

\subsection{Throughput surfaces and joint interpretation}
\begin{figure}[H]
\centering
\subfloat[UDP, 802.11ax, Friis.]{\includegraphics[width=0.475\textwidth]{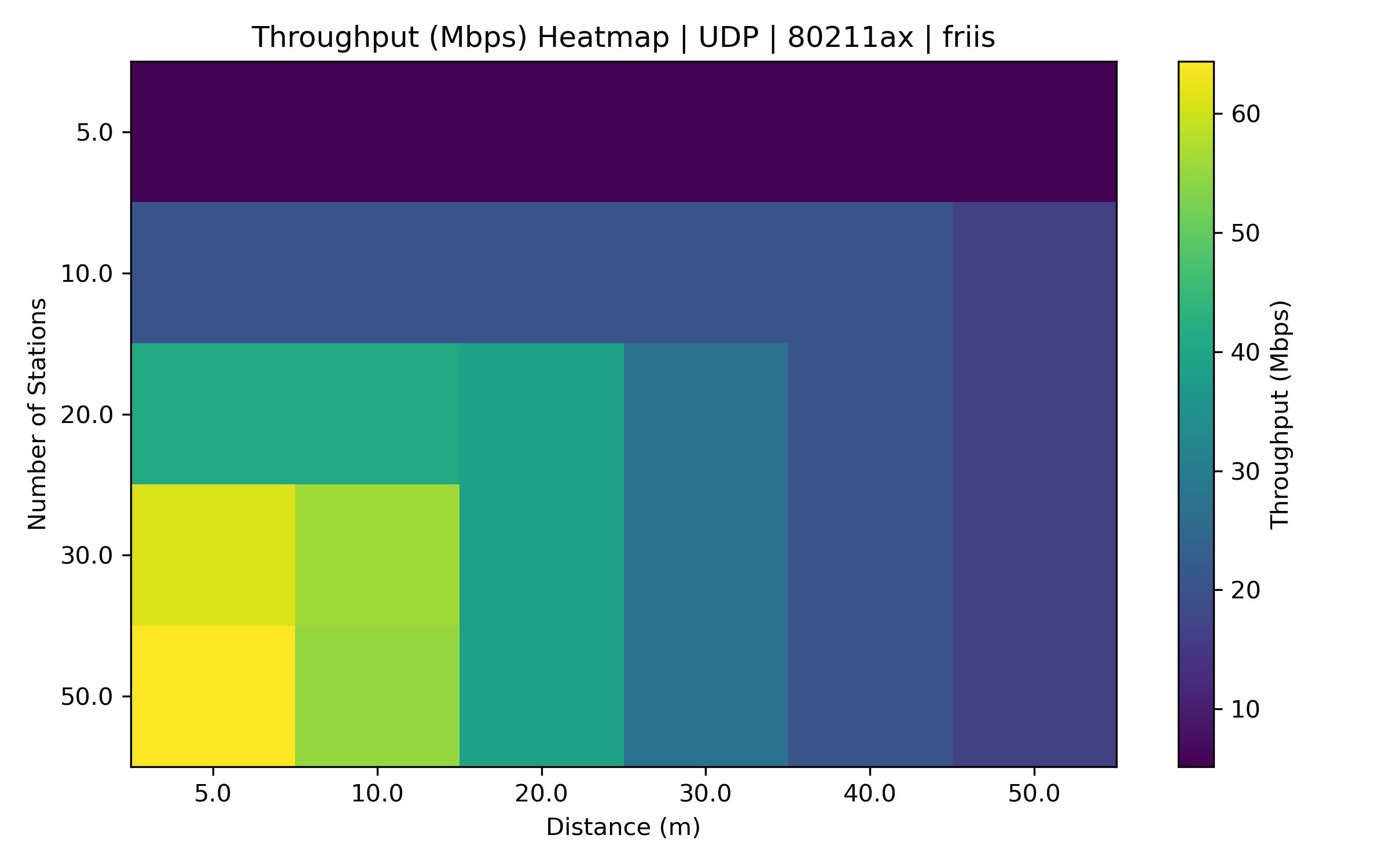}\label{fig:thr-udp-ax-friis}}
\hfill
\subfloat[TCP, 802.11ax, Friis.]{\includegraphics[width=0.475\textwidth]{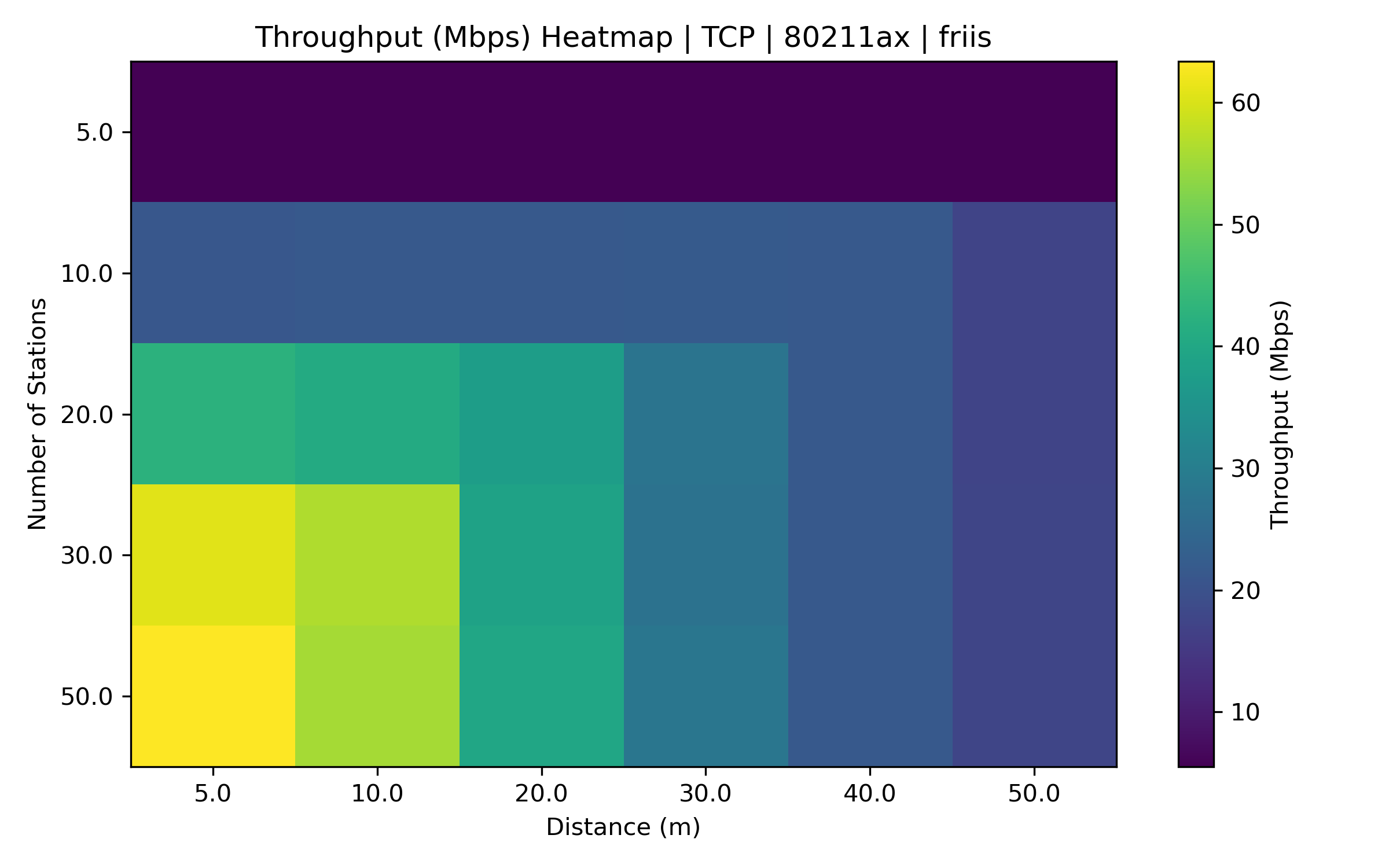}\label{fig:thr-tcp-ax-friis}}\\
\subfloat[UDP, 802.11ax, Friis--Nakagami.]{\includegraphics[width=0.475\textwidth]{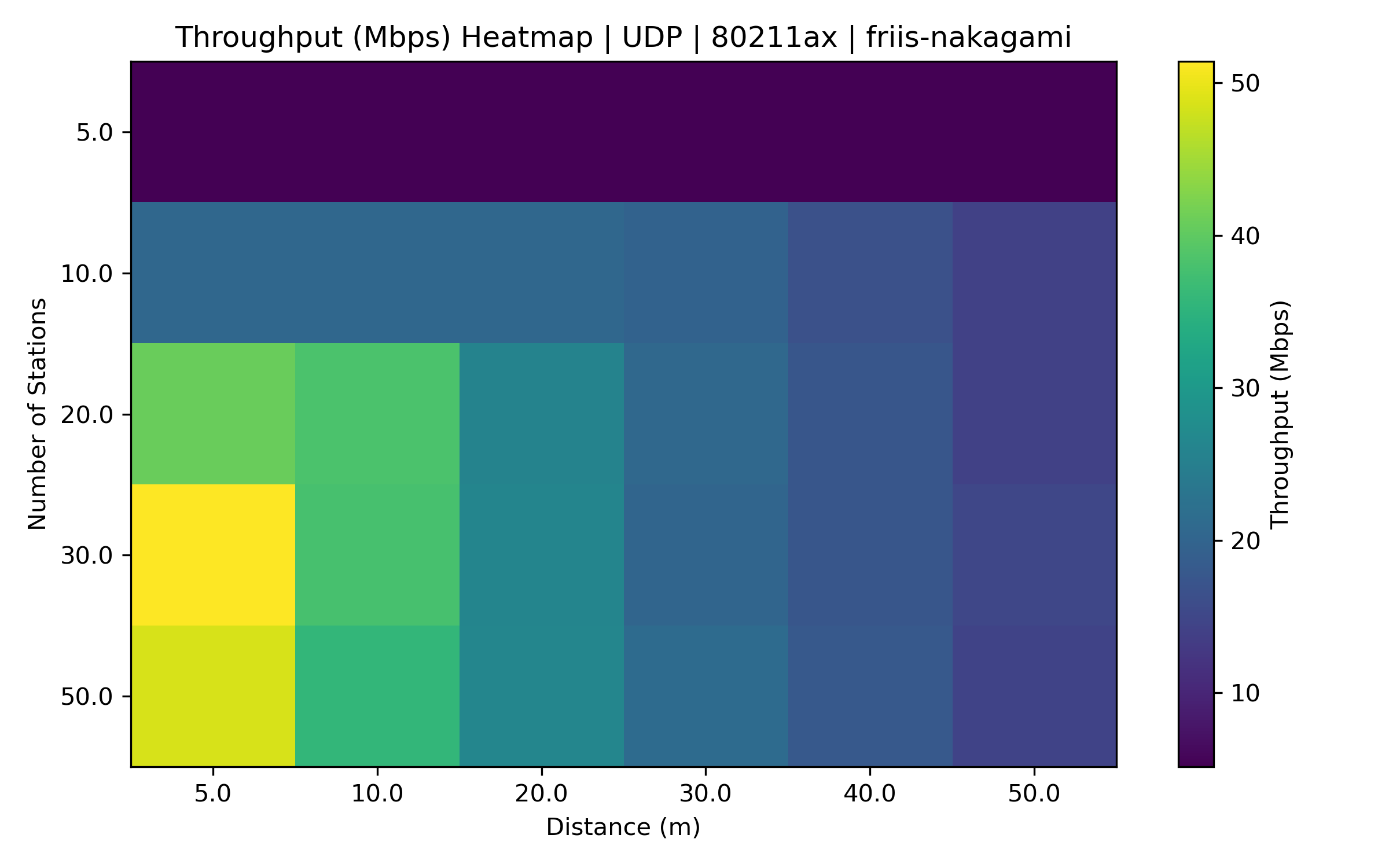}\label{fig:thr-udp-ax-nak}}
\hfill
\subfloat[TCP, 802.11ax, Friis--Nakagami.]{\includegraphics[width=0.475\textwidth]{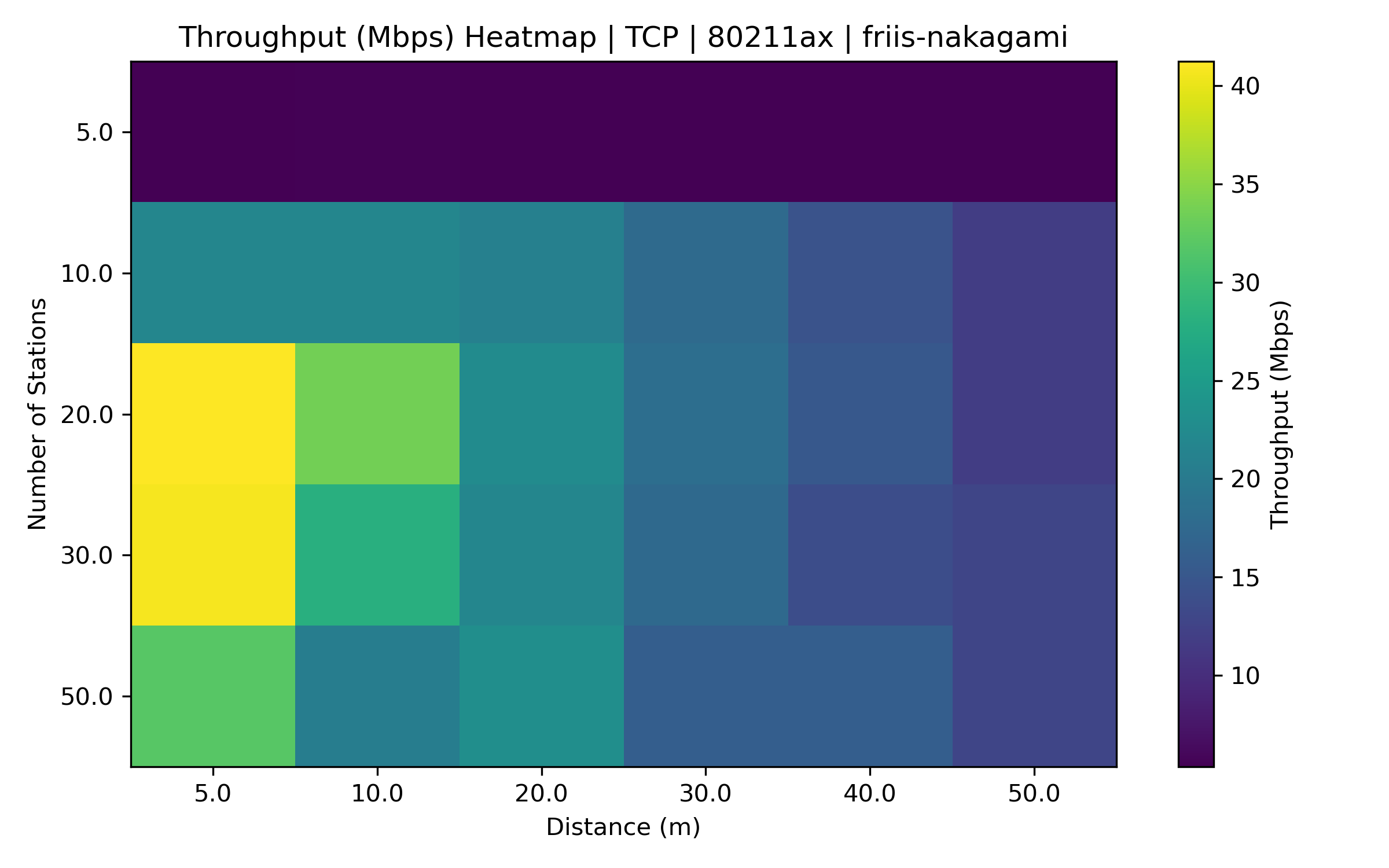}\label{fig:thr-tcp-ax-nak}}
\caption{Selected aggregate-throughput heatmaps over station count and configured distance. Panels retain their original color scales.}
\label{fig:throughputheatmaps}
\end{figure}

In the throughput heatmaps, aggregate throughput can rise as stations are added because additional sources contribute traffic, then saturate as contention increases. Per-station throughput does not follow the same pattern because aggregate capacity is shared among more stations. Figure~\ref{fig:throughputheatmaps} shows selected favourable Friis and fading Friis--Nakagami cases for the configured 802.11ax mode. The heatmaps show lower TCP throughput in several dense fading cells, consistent with TCP responding to loss and delay while UDP continues at its configured source rate.

Taken together, the results answer RQ1 by showing that propagation and the density--distance condition matter jointly: degraded conditions reduce useful throughput and UDP packet delivery, while service inequality can remain even when aggregate throughput is non-trivial. RQ2 is answered by the large gap between TCP and UDP PDR and by cases where TCP throughput does not exceed UDP throughput. For RQ3, the configured 802.11ax mode attains the highest global throughput under Friis, but 802.11ax does not dominate every transport--propagation metric combination. The results do not measure the contribution of OFDMA, MU-MIMO, channel width, or a specific MCS in isolation.

% Keep all Results figures within the Results section instead of allowing
% deferred floats to move past the Discussion or Conclusion.
\FloatBarrier
\section{Discussion}
\subsection{Propagation and density}
The observed Friis results provide an optimistic reference, while the LogDistance and Friis--Nakagami configurations display lower global capacity and/or fairness in several combinations. This is physically plausible because distance-dependent attenuation and fading can reduce received power and make frame delivery less reliable. However, the experimental record does not specify the carrier frequency, reference distance, path-loss exponent, Nakagami parameters, transmit power, antenna gains, or channel configuration. Consequently, the observed ordering and heatmap patterns are reported without attributing numerical differences to a reconstructed link budget.

Increasing station count changes both the nominal offered load and contention. The experiment uses a per-source rate of 1\,Mb/s for five stations and 2\,Mb/s for 10--50 stations, so density is not varied at constant aggregate offered load. Aggregate throughput trends must be read alongside per-station throughput, PDR, delay, fairness, and the configured load. A design that holds aggregate offered load constant would answer a different research question and was not included in this campaign.

\subsection{Transport and interpretation of metrics}
The high TCP PDR is consistent with transport feedback and recovery, whereas UDP exposes losses without transport-layer retransmission. Nevertheless, TCP does not restore radio capacity: retransmissions, acknowledgements, and congestion response consume resources or reduce sending pressure. TCP/UDP WLAN interactions have been documented in prior work \cite{xylomenos1999,bruno2008}. These results should be interpreted conservatively because FlowMonitor counters are IP-flow measurements and reverse ACK flows were excluded from useful-throughput aggregation. Neither these data nor the plotted PDR directly reveal unique application bytes delivered.

Delay also requires joint interpretation. A very small mean delay can coexist with low PDR, low throughput, and low fairness, particularly when the statistic is conditioned on packets counted as received. Such values do not establish that the corresponding operating condition is preferable. Likewise, $\eta_{\mathrm{rel}}$ is not spectral efficiency and may exceed one under the source's byte-accounting and offered-load definitions.

\section{Limitations and Reproducibility}
The campaign is a simulation study with one AP, static stations, uplink traffic, a 10\,s run time, and no reported mobility, obstacles, co-channel external interference, or multiple BSSs. Only three seeds were scheduled, and seven runs failed; affected cells therefore have fewer replications. Confidence intervals based on two or three replications are descriptive and should not be treated as strong evidence of statistical significance \cite{law1977}.

Several implementation details needed for exact independent reproduction were not documented: the ns-3 release/commit; channel number, carrier frequency and width; transmit power and antenna gains; precise STA coordinates and distance-placement rule; LogDistance reference distance and exponent; Nakagami $m$/$\Omega$ parameters; queue and retry limits; TCP variant; and the exact endpoint definition for each FlowMonitor observation interval. The simulation and analysis scripts and run-level raw CSVs were unavailable for this study. These materials are required for exact independent reproduction.

The experiment also does not report application-level sequence-number goodput or packet delivery, direct PHY error counters, or isolated 802.11ax multi-user scheduling. No hypothesis tests or new confidence intervals are added here. Aggregate CSVs preserve the scenario summaries, seed counts, and \texttt{ci95} fields for follow-up analysis.

\section{Conclusion}
This study evaluates how density, distance, propagation, Wi-Fi configuration, and transport protocol jointly relate to WLAN performance. The factorial design scheduled 1,080 runs across UDP/TCP, two Wi-Fi modes, three propagation configurations, five station counts, six distance settings, and three seeds; 1,073 valid CSV outputs were available for aggregation. TCP FlowMonitor PDR was high across the global summaries, but this did not imply greater radio capacity or fairness. The highest reported mean aggregate throughput occurred for configured 802.11ax under Friis, while UDP exceeded TCP under 802.11ax Friis--Nakagami. LogDistance cases illustrate why near-unity TCP PDR and low measured delay cannot be interpreted in isolation from throughput and fairness.

The findings support a multi-metric interpretation of WLAN behaviour and a cautious reading of transport-layer measurements. A complete reproducibility record should include the exact simulator, radio, propagation, TCP, and flow-timing parameters. Future work should examine additional random seeds, mobility, multiple APs, interference, channel-width sweeps, explicit 802.11ax OFDMA/MU-MIMO configurations, and alternative TCP congestion-control algorithms.

\section*{Data and Code Availability}
The project repository provides experiment orchestration and analysis scripts, configuration matrices, processed scenario summaries, and figures (\href{https://github.com/Silima1/ns3-wifi-density-propagation-study}{repository on GitHub}). Run-level raw CSV outputs and the ns-3 simulation source are not included in the repository, so independent reruns of the full simulation campaign require those materials.

\bibliographystyle{IEEEtran}
\bibliography{references}

\end{document}